\documentclass[twocolumn,twocolappendix]{aastex63}

\received{}
\revised{}
\accepted{}

\submitjournal{The Astrophysical Journal}

\shorttitle{Gamma-Rays and Neutrinos from Starburst Galaxies}
\shortauthors{Ha, Ryu, \& Kang}

\begin{document}

\title{Modeling of Cosmic-Ray Production and Transport and Estimation of Gamma-Ray and Neutrino Emissions in Starburst Galaxies}

\author[0000-0001-7670-4897]{Ji-Hoon Ha}
\affil{Department of Physics, School of Natural Sciences, UNIST, Ulsan 44919, Korea}
\author[0000-0002-5455-2957]{Dongsu Ryu}
\affil{Department of Physics, School of Natural Sciences, UNIST, Ulsan 44919, Korea}
\author[0000-0002-4674-5687]{Hyesung Kang}
\affil{Department of Earth Sciences, Pusan National University, Busan 46241, Korea}

\correspondingauthor{Ji-Hoon Ha}
\email{hjhspace223@unist.ac.kr}

\begin{abstract}


Starburst galaxies (SBGs) with copious massive stars and supernova (SN) explosions are the sites of active cosmic-ray production. Based on the predictions of nonlinear diffusive shock acceleration theory, we model the cosmic-ray proton (CRP) production by both pre-SN stellar winds (SWs) and supernova remnants (SNRs) from core-collapse SNe inside the starburst nucleus. Adopting different models for the transport of CRPs, we estimate the $\gamma$-ray and neutrino emissions due to $pp$ collisions from nearby SBGs such as M82, NGC253, and Arp220. We find that with the current $\gamma$-rays observations by Fermi-LAT, Veritas, and H.E.S.S., it would be difficult to constrain CRP production and transport models. Yet, the observations are better reproduced with (1) the combination of the single power-law (PL) momentum distribution for SNR-produced CRPs and the diffusion model in which the CRP diffusion is mediated by the strong Kolmogorov-type turbulence of $\delta B/B\sim1$, and (2) the combination of the double PL model for SNR-produced CRPs and the diffusion model in which the scattering of CRPs is controlled mostly by self-excited waves rather than the pre-existing turbulence. The contribution of SW-produced CRPs could be substantial in Arp220, where the star formation rate is higher and the slope of the initial mass function would be flatter. We suggest that M82 and NGC253 might be detectable as point sources of high-energy neutrinos in the upcoming KM3NET and IceCube-Gen2, when optimistic models are applied. Future observations of neutrinos as well as $\gamma$-rays would provide constraints for the production and diffusion of CRPs in SBGs.

\end{abstract}

\keywords{cosmic rays -- galaxies: starburst -- gamma rays -- neutrinos}

\section{Introduction} 
\label{sec:s1}

A starburst galaxy (SBG) is characterized by a nuclear region, named as the starburst nucleus (SBN), with enhanced star formation rate (SFR) \citep[e.g.,][and references therein]{kennicutt2012}. 
Young massive stars blow powerful stellar winds (SWs) throughout their lifetime and die in spectacular explosions as core-collapse supernovae (SNe) \citep{weaver1977,freyer2003,georgy2013,janka2012}.
Various shock waves driven by these activities can accelerate cosmic-rays (CRs) primarily through diffusive shock acceleration (DSA) \citep[e.g.,][]{drury1983}. 
Some of mechanical energy of SWs is expected to be converted to CRs at
termination shocks in stellar wind bubbles and superbubbles created by young star clusters, 
colliding-wind shocks in binaries of massive stars, and bow shocks of massive runaway stars \citep[e.g.,][]{casse1980,cesarsky1983,bykov2014}.
\citet{seo2018}, for instance, estimated that the energy deposited by SWs could reach up to $\sim 25 \%$ of that from SNe in our Galaxy. In SBGs, the relative contribution of SWs could be more important, since the the integrated galactic
initial mass function (IGIMF) seems to be in general flatter for higher SFR \citep[e.g.,][]{palla2020, weidner2013, yan2017}.
Supernova remnant (SNR) shocks driven by core-collapse SNe propagate into the pre-SN winds with
$\rho\propto r^{-2}$ density profile and strong magnetic fields, 
so the blast waves do not decelerate significantly and
the maximum energy of CR nuclei accelerated by such shocks could reach up to $\sim100$ PeV 
\citep[e.g.,][]{berezhko2000,zirakashvili2018}. Hence, the SBNs of SBGs have been recognized as sites of very active CR production \citep[e.g.,][]{peretti2019,bykov2020}.

\begin{figure}[t]
\vskip -0. cm
\hskip -0.0 cm
\centerline{\includegraphics[width=0.46\textwidth]{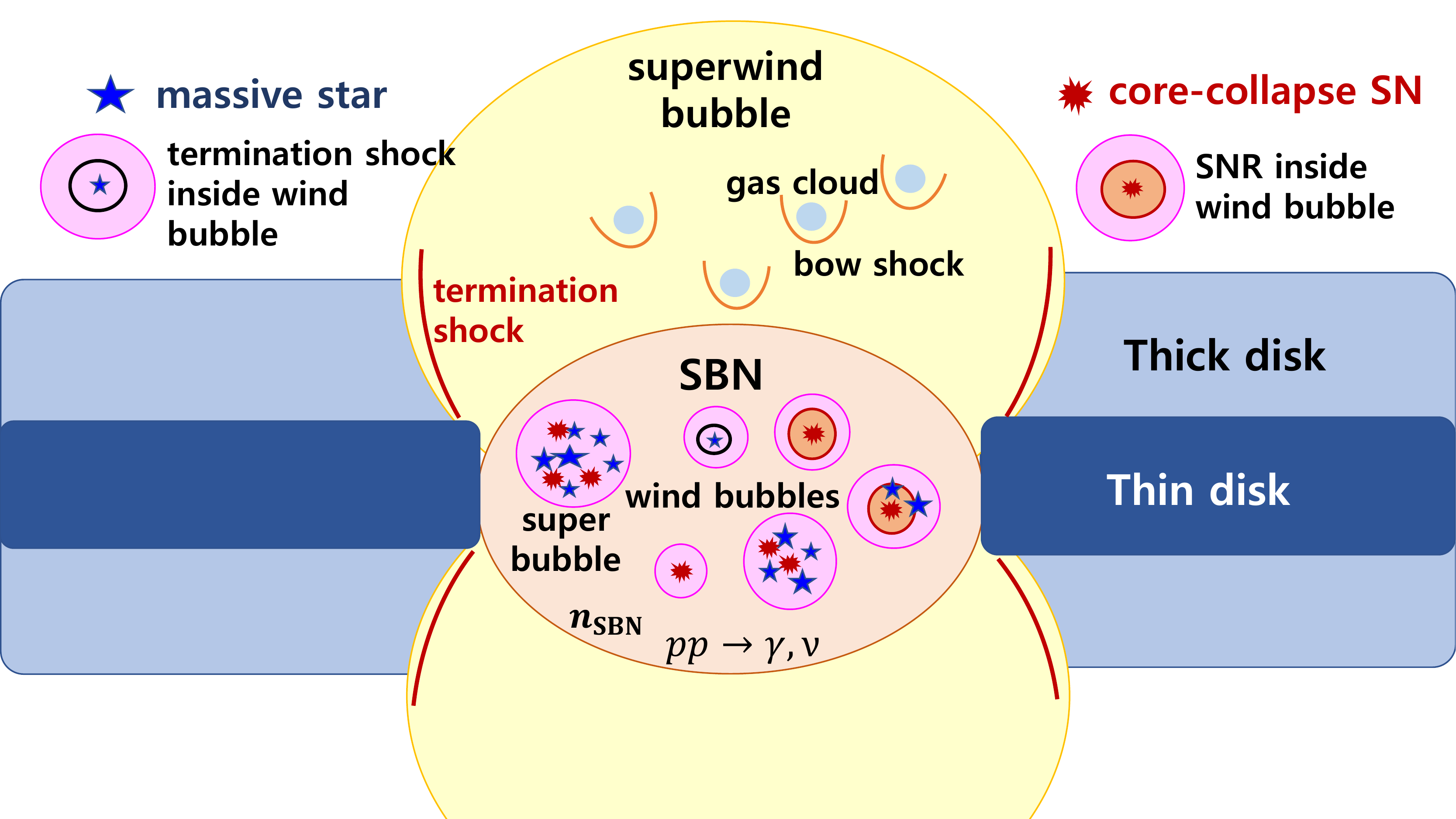}}
\vskip 0.3 cm
\caption{Schematic diagram illustrating a variety of shocks in the central region of SBGs: (1) termination shocks inside stellar wind bubbles or superbubbles created by star clusters \citep{freyer2003}, (2) blast waves from core-collapse SNe inside stellar wind bubbles or superbubbles \citep{berezhko2000}, (3) the termination shock of superwind bubble created by the supersonic outflow from the SBN \citep{romero2018}, and (4) bow shocks around gas clouds colliding with the superwind from the SBN \citep{muller2020}. This study focuses on the CRP production inside the SBN and the ensuing $\gamma$-ray and neutrino emissions due to inelastic $pp$-collisions.
 \label{fig:f1}}
\end{figure}

In addition, the collective input of massive SWs and SN explosions can induce supersonic outflows in SBGs, known as SBN winds, and 
create superwind bubbles around the SBN. It has been proposed that the termination shocks of the superwind and bow shocks formed around gas clouds
embedded in the outflows can provide additional sources of the CR acceleration \citep{romero2018,muller2020}.
In this study, we focus on the CR acceleration at shocks induced by SWs and SNRs inside the SBN (see Figure \ref{fig:f1}).

The beauty of the original DSA predictions lies in the simple power-law (PL) momentum distribution, $f_p(p)\propto p^{-q}$, in the
test-particle regime, where the PL slope, $q = 4M_{\rm s}^2/(M_{\rm s}^2-1)$, depends only on the shock sonic Mach number,
$M_{\rm s}$ \citep[e.g.,][and see further discussions in Section \ref{sec:s2.1}]{drury1983}.
According to the nonlinear theory of DSA, however, with a conversion efficiency of order of 10 \% from the shock kinetic energy to the CR energy at strong shocks, the dynamical feedback of CR protons (CRPs) may lead to a substantial concave curvature in their momentum distribution \citep[e.g.,][]{berezhko1999,caprioli2009,kang2012,bykov2014b}. 
On the other hand, the time-integrated, cumulative CRP spectra produced during the SNR evolution seem to exhibit smaller concavities, 
compared to the instantaneous spectra at the shock position at a given age  \citep[e.g.,][]{caprioli2010,
zirakashvili2012,kang2013}.
The caveat here is that the physics of DSA has yet to be fully understood from
the first principles in order to make quantitative predictions \citep[e.g.][]{marcowith2016}. 
In this study, in addition to the canonical single PL form, we adopt double PL momentum spectra to represent 
slightly concave spectra of CRPs that might be produced by an ensemble of SNRs with wide ranges of parameters
at different dynamical stages.

The CRPs in the SBN can inelastically collide with background thermal protons ($pp$ collisions) and produce neutral and charged pions, which decay into $\gamma$-rays and neutrinos, respectively. Indeed, $\gamma$-rays from SBGs have been observed using Fermi-LAT \citep[e.g.,][]{abdo2010,acero2015,peng2016,Abdollahi2020} and ground facilities, such as H.E.S.S. and Veritas \citep[e.g.,][]{veritas2009,acero2009,fleischhack2015,hess2018}, proving the active acceleration of CRPs in SBGs. Although SBGs have not yet been identified as point sources of neutrinos, they have been considered to be potential contributors of high energy neutrinos observable in IceCube \citep[e.g.,][]{loeb2006,murase2013,tamborra2014,bechtol2017,ajello2020}. 
Adopting a scaling relation between $\gamma$-ray and infrared luminosities, for instance, \citet{tamborra2014} estimated the $\gamma$-ray background and the diffuse high-energy neutrino flux from star-forming galaxies (SFGs) including SBGs, and suggested that SBGs could be the main sources of IceCube neutrinos. \citet{bechtol2017}, on the other hand, argued that considering the contribution from blazers, SFGs may not be the primary sources of IceCube neutrinos \citep[see also][]{ajello2020}. In addition, \citet{peretti2020} argued that SBGs would be hard to be detected as point neutrino sources in IceCube.

There have been previous studies to explain the observed spectra of SBGs in $\gamma$-rays (as well as in other frequency ranges) and to predict neutrino emissions \citep[e.g.,][for recent works]{wang2018,sudoh2018,peretti2019,peretti2020,krumholz2020}. 
The hadronic process for the production of $\gamma$-rays and neutrinos is governed mainly by the ``production'' and ``transport'' of CRPs. 
In most of these previous studies, CRPs are assumed to be produced at SNR shocks, and the cumulative momentum distribution of CRPs inside the SBN is described with the single PL spectrum ($\propto p^{-\alpha}$); the Fermi-LAT $\gamma$-ray observations for well-studied SBGs, such as M82, NGC253, and Arp220, require the CRP momentum distribution with $\alpha \sim 4.2 - 4.5$.

The transport of CRPs in the SBN is expected to be regulated by the advection due to the superwinds escaping from the SBN and the diffusion mediated by magnetic turbulence.
The wind transport is energy-independent and the advection time-scale of CRPs would be comparable to the energy loss time-scale, $\tau_{\rm loss}(p)$, in SBGs \citep[e.g.,][]{peretti2019}.
On the contrary, the diffusive transport is expected to depend on the CRP energy. 
It has been modeled, for instance, with the large-scale, Kolmogorov-type turbulence of $\delta B/B \sim 1$ expected to be present 
in the interstellar medium (ISM) \citep[e.g.,][]{sudoh2018,peretti2019}, and then for most CRPs, the diffusion time-scale, $\tau_{\rm diff}(p)\sim L^2/D(p)$, is longer than $\tau_{\rm loss}(p)$ (see Figure \ref{fig:f3} below). 
\citet{krumholz2020}, on the other hand, argued that the turbulence of scales less than the gyroradius of CRPs with $E_{\rm CR} \lesssim {\rm several} \times 0.1~{\rm PeV}$ would be wiped out by ion-neutral damping, and suggested that CRPs interact mostly with the Alfv\'en waves that are self-excited via CRP streaming instability. 
With this model, in the energy range of $E_{\rm CR} \gtrsim 1~{\rm TeV}$, $\tau_{\rm diff}(p)\ll \tau_{\rm loss}(p)$, so CRPs are inefficiently confined in the SBN. 

Previous studies have been successful to some degree in modeling the production and transport of CRPs in SBGs and reproducing/predicting $\gamma$-ray and neutrino emissions from them.
In this work, we expand such efforts; we consider different models for CRP production and transport, and estimate $\gamma$-ray and neutrino emissions for nearby SBGs, specifically, M82, NGC253, and Arp220.
First, in addition to the single PL momentum distribution, we consider the double PL distribution for SNR-produced CRPs. 
This is partly motivated by the employment of double PLs in fitting observed $\gamma$-ray spectra \citep[e.g.,][]{tamborra2014,bechtol2017}, but also by the consideration of nonlinear DSA theory as discussed above.
We also consider the CRPs produced at SW shocks.
Second, we adopt different diffusion models for the diffusive transport of CRPs in the SBN, specifically the model where CRPs scatter off the Kolmogorov turbulence of $\delta B/B \sim 1$ and the model suggested by \citet{krumholz2020}. 
We then examine the consequence of different momentum distributions of CRPs, combined with different diffusive transport models in the estimation of $\gamma$-ray and neutrino emission from SBGs. Finally, we discuss the implications of our results.

In this study, we consider the $\gamma$-rays and neutrinos emitted from the SBN only \citep[see also, e.g.,][]{peretti2019,peretti2020}, while ignoring the contribution from the galactic disk. Since both the gas density and CRP density should be substantially lower in the disk than in the SBN, the disk would make relatively a minor contribution.

This paper is organized as follows. In Sections \ref{sec:s2} and \ref{sec:s3}, we detail models for the production and transport of CRPs in the SBN. In Sections \ref{sec:s4} and \ref{sec:s5}, we present the estimates of $\gamma$-rays and neutrinos from SBGs. A summary including the implications follows in Section \ref{sec:s6}.

\section{Cosmic-Ray Production in SBN}
\label{sec:s2}

\subsection{Physics of Nonlinear DSA}
\label{sec:s2.1}

In general, the physics of DSA is governed mainly by both the sonic and Alfv\'en Mach numbers 
and the obliquity angle of magnetic fields in the background flow \citep[e.g.,][]{balogh2013,marcowith2016}.
The outcomes of DSA, in fact, depend on nonlinear interrelationships among complex kinetic processes,
including the injection of thermal particles to Fermi-I acceleration process, 
the magnetic field amplification due to resonant and non-resonant CR streaming instabilities,
the drift of scattering centers relative to the underlying plasma, 
the escape of highest energy particles due to lack of resonant scattering waves, and the precursor heating due to 
wave damping \citep[e.g.,][]{ptuskin2005,vladimirov2008,caprioli2009,zirakashvili2012,kang2012}.

In the test-particle regime, the DSA theory predicts that the CRPs accelerated via Fermi-I process have
the PL momentum spectrum whose slope is determined by the velocity jump of scattering centers across the shock
\citep{bell1978}:
\begin{equation}
\alpha = \frac{3 (u_1 + u_{\rm w1})}{(u_1 + u_{\rm w1})-(u_2 + u_{\rm w2})},
\label{eq1}
\end{equation}
where $u_1$ and $u_2$ are the bulk flow speeds in the shock rest frame,
while $u_{\rm w1}$ and $u_{\rm w2}$ are the mean drift speeds of scattering centers relative to the underlying flows.
Hereafter, we use the subscripts ‘1’ and ‘2’ to denote the preshock and postshock states, respectively.
This slope could be greater than the slope estimated with the flow speeds only, $q= 3u_1/(u_1-u_2)$ (quoted in the introduction), 
since the scattering centers are likely to drift away from the spherically expanding SNR shock in both the preshock and postshock regions with 
speeds similar to the local Alfv\'en speeds. 
For example, $u_{\rm w1}\sim - v_{A1}$ and $u_{\rm w2}\sim + v_{A2}$
(but note that $u_{\rm w2} \approx 0$ for isotropic turbulence in the downstream flow)
\citep[e.g.][]{berezhko2000,kang2010,kang2012}, and then
such Alfv\'enic drift of scattering centers tends to steepen the test-particle spectrum \citep[e.g.][]{kang2018}. 
The drift speed of scattering centers depends on the magnetic fields in the underling plasma, 
while the amplification of turbulent magnetic fields due to CR streaming instabilities depends on the CR 
acceleration efficiency, which in turn is controlled by the injection rate \citep{caprioli2009,kang2013}. 
The full understanding of such nonlinear interdependence among different plasma processes still remains quite challenging.

In the nonlinear DSA theory, on the other hand, the concavity in the CRP spectrum arises due to the development of 
a smooth precursor upstream of the subshock \citep[e.g.,][]{kang2007,caprioli2009}. 
Then, low energy CRPs with shorter diffusion lengths experience smaller velocity jumps across the subshock, 
whereas high energy CRPs with longer diffusion lengths feel larger velocity jumps across the total shock structure.
As a result, the slope becomes $\alpha_{\rm sub} > \alpha$ (steeper) at the low energy part of the CRP spectrum,
while it becomes $\alpha_{\rm tot} < \alpha$ (flatter) at the high energy part.
Here, $\alpha_{\rm sub}$ ($\alpha_{\rm tot}$) is estimated by Equation (\ref{eq1}) with the velocity jump across the subshock (total shock) transition.
Obviously, this nonlinear effect could be substantial for strong SNR shocks with high acceleration efficiencies,
although the predicted efficiency depend on the phenomenological models for various processes adopted in numerical studies 
\citep[e.g.,][]{berezhko1999,vladimirov2008,bykov2014b,caprioli2010,kang2013} and hence it is still not very well determined.
In the case of spherically expanding SNRs, however, the cumulative CRP spectrum integrated during the SNR evolution 
may exhibit only a mild concavity, as mentioned in the introduction. 

Another critical issue is the highest energy of CRPs, $E_{\rm max}= p_{\rm max}c$, accelerated by the blast waves
from core-collapse SNe.
SNRs propagate into slow red-supergiant (RSG) winds in the case of Type II
or fast Wolf-Rayet (WR) winds in the case of Type Ib/c, before they run into the main-sequence (MS) wind shells
\citep{berezhko2000,georgy2013}.
For instance, WR winds are metal-enriched and have strong magnetic fields, 
and the maximum energy of CR nuclei accelerated by such shocks could reach well above PeV.
In addition, the blast waves do not significantly decelerate in the $\rho\propto r^{-2}$ wind flows,
the magnetic fields of SWs are stronger than the mean ISM fields,
and the shock geometry is likely to be quasi-perpendicular due to the Parker-spiral magnetic field lines 
originating from the stellar surface \citep[e.g.,][]{ptuskin2005,tatischeff2009,zirakashvili2016,zirakashvili2018}; all these would affect the achievable $E_{\rm max}$. Considering uncertainties in $E_{\rm max}$, we here adopt the highest momentum of CRPs $p_{\rm max} \simeq 1 - 100$ PeV/$c$, based on previous numerical studies. 

\subsection{Cosmic-Ray Production at Supernova Remnant Shocks}
\label{sec:s2.2}

Taking account of nonlinear effects, we consider double (broken) PL momentum spectra, in addition to canonical single PL spectra, for CRPs
produced by the collective SN explosions inside the SBN. The single PL form for the CRP production rate is given as
\begin{equation}
\mathcal{N}_{\rm SN,p}(p) \propto \left(\frac{p}{m_p c}\right)^{-\alpha_{\rm SN}} {\rm exp}\left(-\frac{p}{p_{\rm max}}\right)
\label{eq2}
\end{equation}
with an exponential cutoff at $p_{\rm max}$. 
Here, the slope $\alpha_{\rm SN}$ and $p_{\rm max}$ are free parameters. 
We adopt $\alpha_{\rm SN} \simeq 4.2 - 4.5$, which is consistent with the range of values previously used for the modeling of $\gamma$-ray observations in the energy range of $E_{\gamma} \lesssim 100$ GeV \citep[e.g.,][]{tamborra2014,peretti2019}.

The double PL form for the CRP production rate is given as
\begin{equation}
\mathcal{N}_{\rm SN,p}(p) \propto \left\{\begin{array}{lr} \left({\displaystyle p\over \displaystyle m_p c}\right)^{-\alpha_{\rm SN,1}}~~~~~~~~~~~~~~~~~~~~(p < p_{\rm brk})\\
\left({\displaystyle p\over \displaystyle m_p c}\right)^{-\alpha_{\rm SN,2}} {\rm exp}\left(-{\displaystyle p\over \displaystyle p_{\rm max}}\right)~~(p \geq p_{\rm brk}), 
\end{array}\right. \nonumber\\ 
\label{eq3}
\end{equation}
where $p_{\rm brk}$ is the break momentum between the two PLs. 
For a typical concave CRP spectrum integrated over the SNR age, 
$\alpha_{\rm SN,2}$ is expected to be only slightly smaller than $\alpha_{\rm SN,1}$, while $p_{\rm brk} \simeq 10^{-3}-10^{-2} p_{\rm max}$ \citep{berezhko2000,tatischeff2009,kang2013,zirakashvili2016}. We choose $\alpha_{\rm SN,1} \simeq 4.2 - 4.5$ and $\alpha_{\rm SN,2} \simeq 3.95- 4.1$, modeling a mild concavity.
Note that $\mathcal{N}_{\rm SN,p}(p)$ represents the CRP spectrum due to an ensemble of SNRs 
from different types of core-collapse SNe at different dynamical stages, and $ \int 4\pi p^2 \mathcal{N}_{\rm SN,p}(p) dp$ gives the volume-integrated CRP production rate
in the entire volume of the SBN.

The normalization of the single and double PL momentum distributions in Equations (\ref{eq2}) and (\ref{eq3}) is fixed by the following condition:
\begin{equation}
\int_{0}^{\infty}4\pi p^2\mathcal{N}_{\rm SN,p}(p) [\sqrt{p^2c^2+m_p^2c^4}-m_pc^2]dp = \eta_{\rm CR}\mathcal{L}_{\rm SN}.
\label{eq4}
\end{equation}
Here, the amount of energy ejected per second by collective SN explosions in the SBN is given as $\mathcal{L}_{\rm SN}
=\mathcal{R}_{\rm SN} \times E_{\rm SN}~[\rm erg ~s^{-1}]$, where $\mathcal{R}_{\rm SN}$ is the SN rate and $E_{\rm SN}$ is the supernova explosion energy. 
We adopt the values of $\mathcal{R}_{\rm SN}$ given by \citet{peretti2019}.
Assuming the mean explosion energy $E_{\rm SN}=10^{51}$ ergs, 
independent of the initial stellar mass, the estimated values of $\mathcal{L}_{\rm SN}$ for three nearby SBGs are listed in Table \ref{tab:t1}. 
We then assume that a constant fraction, $\eta_{\rm CR} = 0.1$, of $E_{\rm SN}$ is converted to CRPs at each SNR \citep[e.g.,][]{caprioli2014}.

\begin{deluxetable}{ccccccccc}[t]
\tablecaption{Parameters of Three Nearby SBGs Used in Modelings \label{tab:t1}}
\tablenum{1}
\tablehead{
\colhead{} & \colhead{~} &
\colhead{M82} & \colhead{~} &
\colhead{NGC253}& \colhead{~} &
\colhead{Arp220}}
\startdata
$z^a$ & & $9 \times 10^{-4}$ & & $8.8 \times 10^{-4}$ & & $1.76 \times 10^{-2}$ \\
$D_{\rm L}$ [Mpc]$^a$& & 3.9  & & 3.8  & & 77  \\
$R_{\rm SBN}$ [pc]$^a$& & 220  & &  150  & & 250  \\
$n_{\rm SBN}$ [${\rm cm}^{-3}$]$^a$ & & 175   & & 250 & & 3500 \\
$\mathcal{R}_{\rm SN}$ [${\rm yr}^{-1}$]$^a$& & 0.05  & & 0.027  & & 2.25  \\
$v_{\rm SBNwind}$ [${\rm km}~{\rm s}^{-1}$]$^a$ & & 600 & & 300 & & 500 \\
$H_{\rm gas}$ [pc]$^b$& & 73  & &  130  & & 75  \\
$B$ [$\mu$G]$^b$ & & 76 & & 50  & & 1200 \\
$M_{\rm A,turb}^b$ & & 2 & & 2  & & 2 \\
$v_{\rm \rm Ai}$ [${\rm km}~{\rm s}^{-1}$]$^b$ & & 880  & & 920 & & $3500/\sqrt{10}$ \\
$\mathcal{L}_{\rm SN}$ [$10^{40} {\rm erg}~{\rm s}^{-1}$]$^c$ & & 159  & &  85.6  & & 7140  \\
$\mathcal{L}_{\rm SW}$ [$10^{40} {\rm erg}~{\rm s}^{-1}$]$^c$ & & 49  & & 26.2  & & 6200, 3070  \\
\enddata
\tablenotetext{a}{$z$ and $D_{\rm L}$ are the redshift and the luminosity distance of SBGs. $R_{\rm SBN}$ is the radius of the SBN. $n_{\rm SBN}$ and $\mathcal{R}_{\rm SN}$ are the gas number density and the SN rate in the SBN. $v_{\rm SBNwind}$ is the escaping speed of the SBN superwind. The values are from \citet{peretti2019}.}
\tablenotetext{b}{$H_{\rm gas}$ is the gas scale height. $B$, $M_{\rm A,turb}$ and $v_{\rm Ai}$ are the magnetic field strength, the Alfv\'en Mach number of turbulence, and the ion Alfv\'en speed, respectively. The values are from \citet{krumholz2020}.}
\tablenotetext{c}{$\mathcal{L}_{\rm SN}$ and $\mathcal{L}_{\rm SW}$ are the amounts of energy ejected per second by SN explosions and SWs in the SBN. $\mathcal{L}_{\rm SN}$ is estimated using $\mathcal{R}_{\rm SN}$. $\mathcal{L}_{\rm SW}$ is estimated by us (see the text). For Arp220, the larger value of $\mathcal{L}_{\rm SW}$ is calculated with the flatter IGIMF of $\beta = 1$, while the smaller value is obtained with the softer IGIMF of $\beta = 2$.}
\vspace{-0.8cm}
\end{deluxetable}

\subsection{Mechanical Energy Deposition by Stellar Winds}
\label{sec:s2.3}

The current understanding of stellar evolution indicates that massive stars with the initial mass of $12  M_{\sun} \lesssim M_{\rm ZAMS}\lesssim 35 M_{\sun}$
eject the MS wind, followed by the RSG wind, before they explode as Type II SNe. (Here, ZAMS stands for zero-age main sequence.)
More massive stars with $ M_{\rm ZAMS}\gtrsim 35 M_{\sun}$ expel the MS wind, followed by the WR wind,
before they explode as Type Ib/c SN \citep{georgy2013}.
Adopting the IGIMF for our Galaxy and the SN explosion rate of $\mathcal{R}_{\rm SN}=0.015 {\rm yr^{-1}}$, 
\citet{seo2018} estimated that the total wind luminosity emitted by all massive stars in the Galactic disk
is about ${\mathcal L}_{\rm SW} \approx 1/4 {\mathcal L}_{\rm SN}$ in the Milky Way.

To estimate the galaxy-wide wind luminosity, ${\mathcal L}_{\rm SW}$, in the SBN of three nearby SBGs, listed in Table \ref{tab:t1},
we follow the prescription given in \citet{seo2018}, which is briefly described here without repeating all the details.
The first step is to estimate the IGIMF, $N(m) = A_{\rm OB}~m^{-\beta}$, of all stars with the initial mass, $m$,
contained in the SBN, where the normalization factor 
$A_{\rm OB}$ is determined by $\mathcal{R}_{\rm SN}$ of each SBG and the slope $\beta$ depends on the SFR \citep[e.g.,][]{palla2020, weidner2013, yan2017}. For M82 and NGC253 where the SFR is $\lesssim 10 M_{\sun}~{\rm yr}^{-1}$, $\beta = 2.35$ is used, which is consistent with the canonical $N(m)$ by \citet{salpeter1955}. Arp220, on the other hand, has a higher SFR of $\gtrsim 100 M_{\sun}~{\rm yr}^{-1}$, and $N(m)$ is expected to be flatter with $\beta \simeq 1 - 2$ \citep[see also][]{dabringhausen2012}. Hence, we consider $\beta = 1$ and 2 for Arp220.
We then estimate the mass function, $N_k(m)$, in the $k$-th wind stage, which represents MS, RSG, or WR,
using Equation (32) of \citet{seo2018}.
The wind mechanical luminosity at each wind stage, $L_k(M)=1/2 \dot{M}_k v_{{\rm SW},k}^2$, are calculated using their Equations (24)-(28),
where $M$ is the stellar mass at a given stage.
The fitting forms for the mass loss rate, $\dot{M}_k$, and the wind velocity, $v_{{\rm SW},k}$, 
derived from observational and theoretical estimations, are given in \citet{seo2018}.
Note that the conversion of $L_k(M)$ to the wind luminosity as a function of the initial mass $m$, $L_k(m)$, 
requires the estimation of $M$ through numerical calculations of stellar evolutionary tracks \citep[e.g.,][]{ekstrom2012,georgy2013}.
Finally, the galaxy-wide wind luminosity is calculated as
\begin{equation}
\mathcal{L}_{\rm SW} = \sum _k \int N_k(m)L_k(m)dm,
\label{eq5}
\end{equation}
where again $k$ is for the three wind stages,
and the mass range (in units of $M_{\sun}$) of each integral is $[10,150]$ for the MS stage, 
$[10,40]$ for the RSG stage, and $[25,150]$ for the WR stage.

Generally, in galaxies, the SN luminosity, $\mathcal{L}_{\rm SN}$, is larger than the SW luminosity, $\mathcal{L}_{\rm SW}$; here, while $\mathcal{L}_{\rm SW}/\mathcal{L}_{\rm SN} \sim 0.3$ for M82 and NGC253, $\mathcal{L}_{\rm SW}/\mathcal{L}_{\rm SN} \sim 0.87~(\beta = 1) - 0.43~(\beta = 2)$ for Arp220 (see Table \ref{tab:t1}).
In our model, the relative importance of $\mathcal{L}_{\rm SN}$ and $\mathcal{L}_{\rm SW}$ depends on the slope of the IGIMF. This is because while the SN explosion energy ($\sim 10^{51}$ ergs) weakly depends on the pre-SN stellar mass \citep[e.g.,][]{fryer2012}, the mechanical energy of SWs depends rather strongly on the stellar mass \citep[e.g.,][]{seo2018}. We normalize the IGIMF with the SN rate, as mentioned above; then, for a fixed SN rate and energy ($E_{\rm SN} = 10^{51}$ ergs, here), $\mathcal{L}_{\rm SW}/\mathcal{L}_{\rm SN}$ depends on the slope of the IGIMF. Arp220 with a higher SFR has a flatter IGIMF, and hence a higher $\mathcal{L}_{\rm SW}/\mathcal{L}_{\rm SN}$.

\begin{figure*}[t]
\vskip -0.5 cm
\hskip -0.1 cm
\centerline{\includegraphics[width=1.2\textwidth]{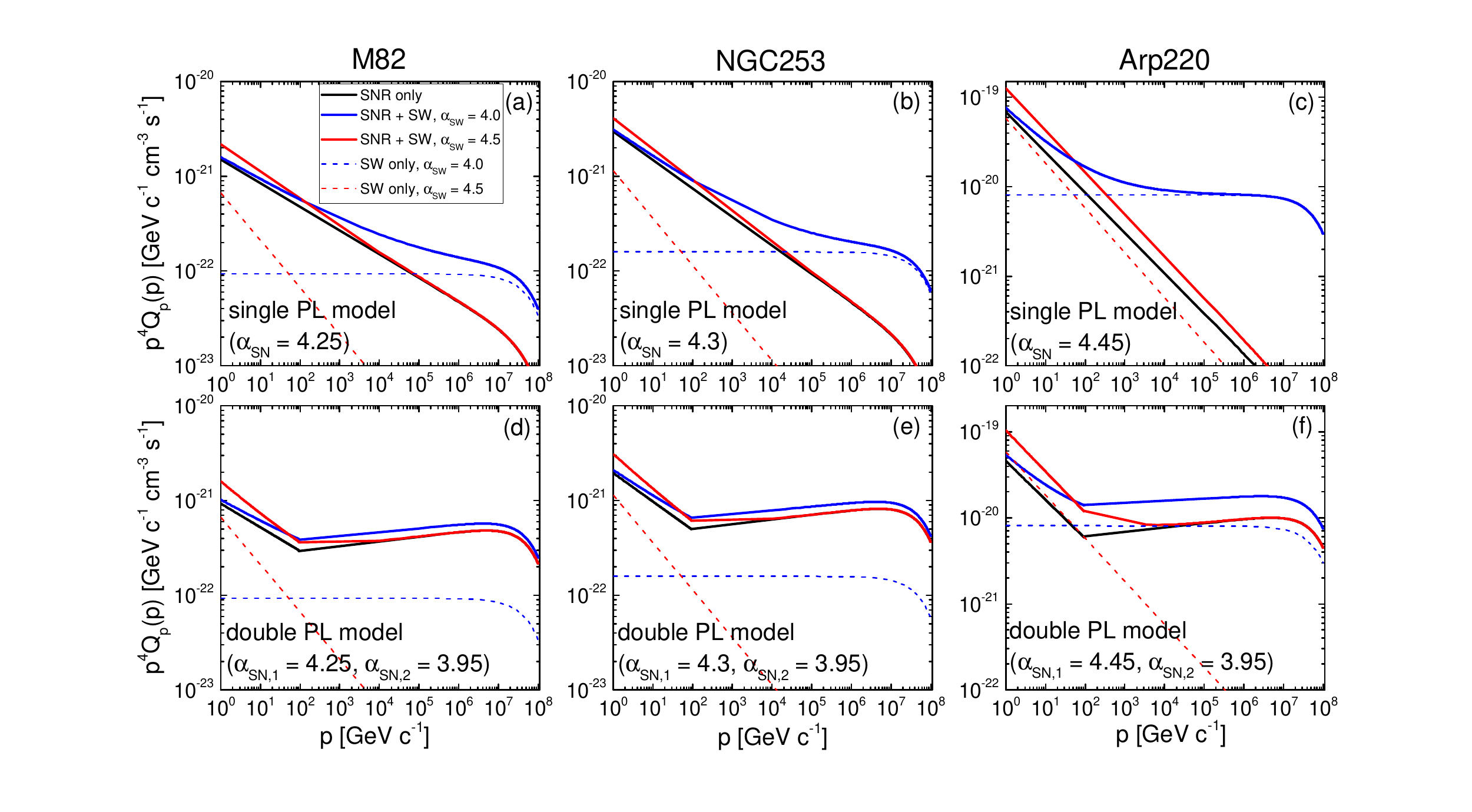}}
\vskip -1. cm
\caption{Injection rate of CRPs, $Q_p(p)$, as a function of the CRP momentum, in the SBN for three nearby SBGs. The black solid lines show $Q_p(p)$ of the SNR-produced CRPs only, which are modeled with either single PL distributions (top panels) or double PL distributions with $p_{\rm brk} = 10^2~m_p c$ (bottom panels). The adopted PL slopes are given in each panel. 
The red and blue dashed lines show the SW-produced CRPs, which are modeled with single PL distributions. Two different PL slopes, $\alpha_{\rm sw}=4.0$ and 4.5, are considered. The red and blue solid lines show $Q_p(p)$ that includes both the SNR and SW-produced CRPs. For Arp220, the cases of $\beta = 1$ are shown.
In all model distributions, $p_{\rm max}=100$ PeV/$c$ is used.\label{fig:f2}}
\end{figure*}

\subsection{Cosmic-Ray Production at Stellar Wind Shocks}
\label{sec:s2.4}

Pre-SN winds are known to induce several kinds of shocks, including termination shocks in the wind flow, forward shocks around 
the wind bubble, colliding-wind shocks in massive binaries, and bow-shocks of massive runaway stars.
Most of the mechanical energy of fast stellar winds is expected to be dissipated at strong termination shocks
during the MS and WR stages.
For typical SWs of O-type MS and WR stars, the wind termination velocity ranges $v_{\rm SW} \sim 1-3 \times 10^3 {\rm km~s^{-1}}$,
and the temperature of the wind flow ranges $T_{\rm SW} \sim 10^4 - 10^5$~K \citep{georgy2013},
so the termination shock is expected to have high Mach numbers and the nonlinear DSA might be important there.
However, the details of DSA physics at SW shocks are rather poorly understood, compared to those at SNRs.
For example, the Alfv\'enic drift effects due to amplified turbulent magnetic fields are somewhat uncertain,
because the magnetic fields are inferred to be strong with quasi-perpendicular configuration at the termination shock.
Moreover, the quantitative calculation of nonlinear DSA at such termination shocks or forward shocks around the wind bubble has not been made so far, unlike in the case of SNRs.

Considering the lack of clear understanding of nonlinear DSA physics and also the CRP contributions from several types of shocks associated with SWs,
for the sake of simplicity, we assume that the cumulative, time-integrated spectra of CRPs produced by SWs 
could be represented by the single PL form,
and that the DSA efficiency is similar to the widely adopted value for SNRs, $\eta_{\rm CR} = 0.1$.
Hence, the CRP production rate due to the collection of shocks induced by SWs from an ensemble of massive stars in the SBN 
is modeled as
\begin{equation}
\mathcal{N}_{\rm SW,p}(p) \propto \left(\frac{p}{m_p c}\right)^{-\alpha_{\rm SW}} {\rm exp}\left(-\frac{p}{p_{\rm max}}\right),
\label{eq6}
\end{equation}
where we adopt $\alpha_{\rm SW} \simeq 4.0 - 4.5$ and $p_{\rm max} \sim 1 - 100$ PeV/$c$, similarly as in $\mathcal{N}_{\rm SN,p}(p)$.
Again, the normalization for $\mathcal{N}_{\rm SW,p}$ for each SBG is fixed by the following relation
\begin{equation}
\int_{0}^{\infty}4\pi p^2\mathcal{N}_{\rm SW,p}(p) [\sqrt{p^2c^2+m_p^2c^4}-m_p c^2]dp = \eta_{\rm CR}\mathcal{L}_{\rm SW},
\label{eq7}
\end{equation}
with $\mathcal{L}_{\rm SW}$ in Table \ref{tab:t1}.

\begin{figure}[t]
\vskip 0 cm
\hskip -0.2 cm
\centerline{\includegraphics[width=0.52\textwidth]{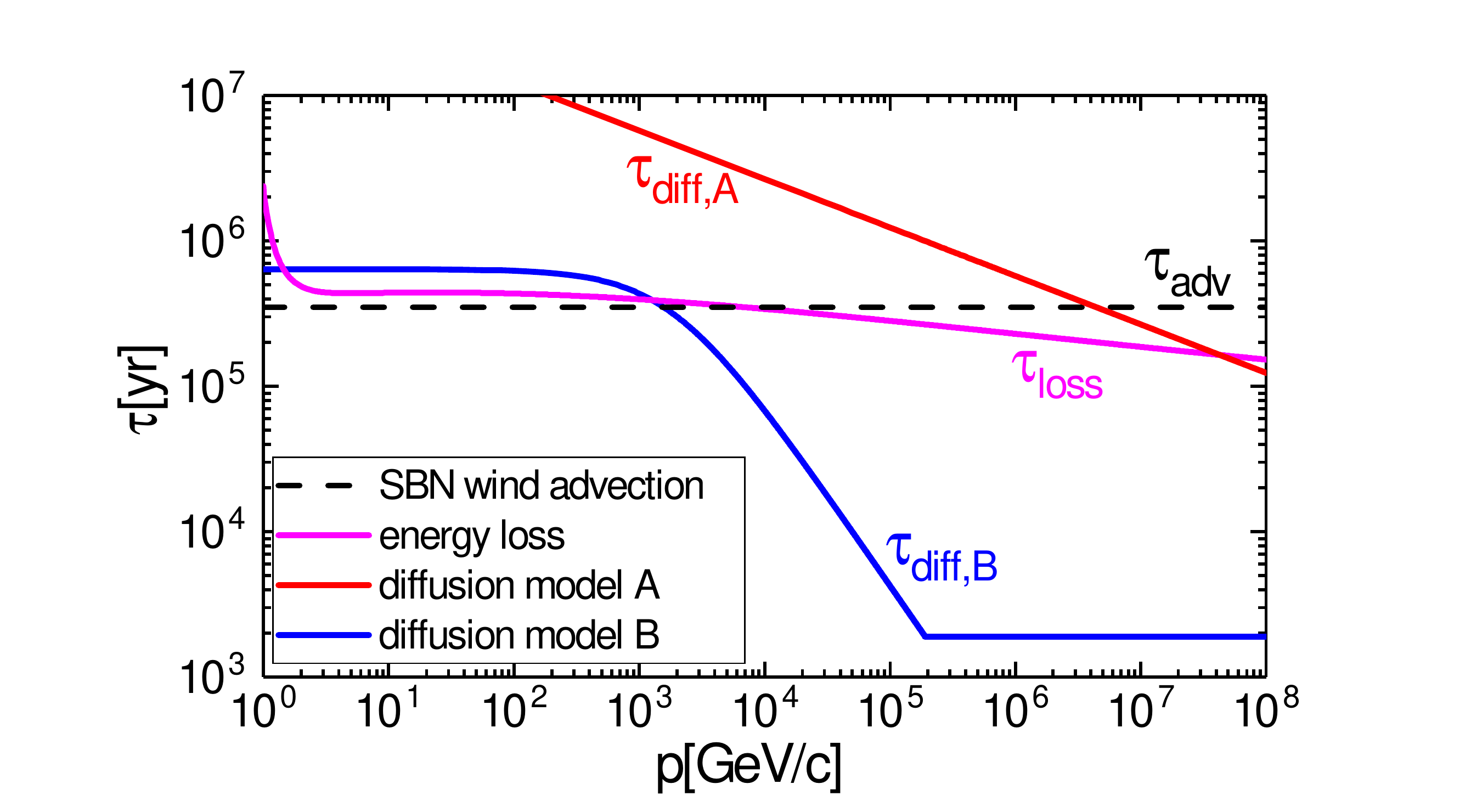}}
\vskip -0.3 cm
\caption{Time-scales in the modeling of CRP transport due to the energy loss, $\tau_{\rm loss}$, SBN wind advection, $\tau_{\rm adv}$, and turbulent diffusion, $\tau_{\rm diff,A}$ and $\tau_{\rm diff,B}$, for M82. In the diffusion model A, CRPs are assumed to scatter off large-scale, Kolmogorov turbulence. 
In the diffusion model B, CRPs interact with self-excited Alfv\'en waves. The line of the model B is for the single PL momentum distribution of SNR-produced CRPs only with $\alpha_{\rm SN}=4.25$ (the black line in Figure \ref{fig:f2}(a)), and the kink around $p\sim0.1$ PeV/$c$ is caused by the limit of streaming speed, $v_{\rm st} < c$. \label{fig:f3}}
\end{figure}

\begin{figure*}[t]
\vskip -0.4 cm
\hskip -0.4 cm
\centerline{\includegraphics[width=1.2\textwidth]{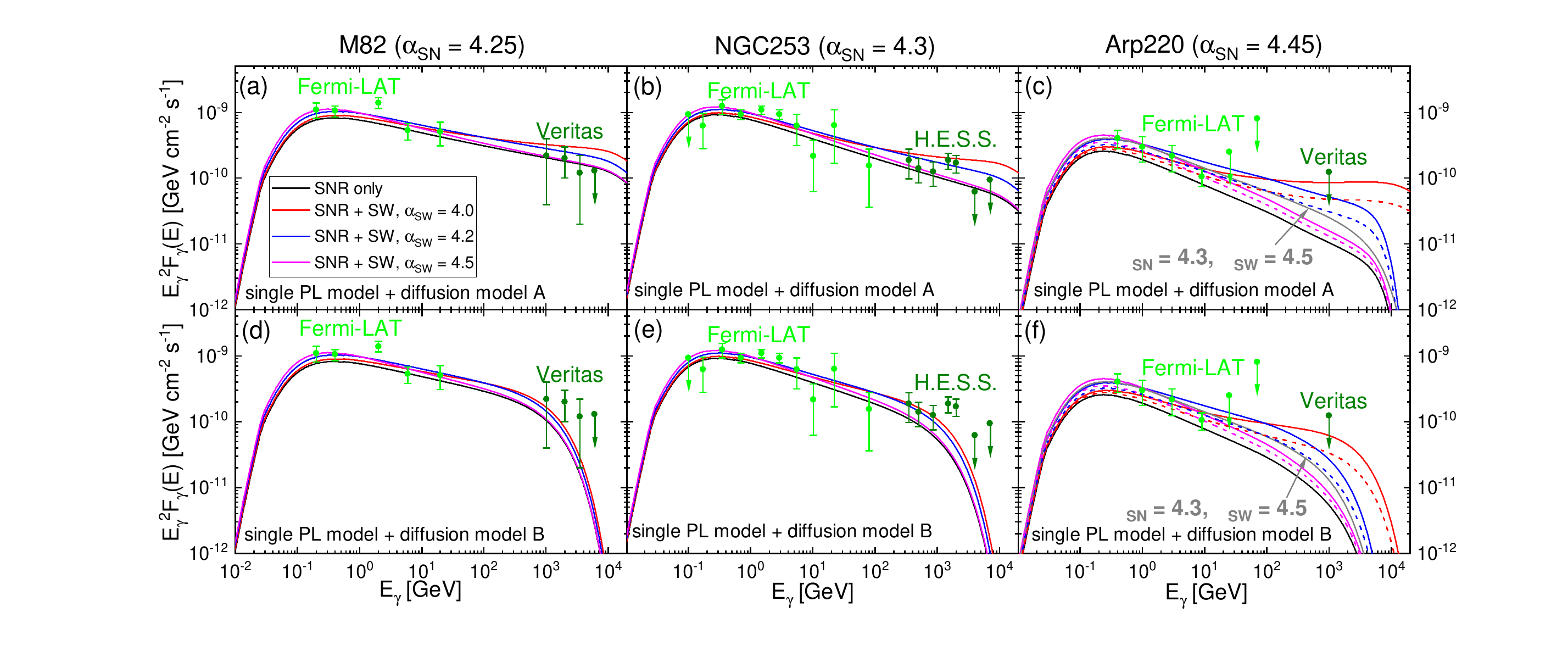}}
\vskip -0.5 cm
\caption{The $\gamma$-ray fluxes, $F_{\gamma}$, estimated with the single PL distributions for both SNR-produced CRPs and SW-produced CRPs, for M82 in panels (a) and (d), NGC253 in panels (b) and (e), and Arp220 in panels (c) and (f). For the slope of the SNR-produced CRP distribution $\alpha_{\rm SN}$, the best fitted values quoted in \citet{peretti2019} are used, while for the slope of the SW-produced CRP distribution $\alpha_{\rm SW}$, a number of different values are tried. The upper panels show $F_{\gamma}$ with the diffusion model A, while the bottom panels show $F_{\gamma}$ with the diffusion model B. For Arp220, the cases of $\beta = 1$ and $\beta = 2$ are shown with the solid and dashed lines, respectively; in addition, the gray solid lines plot $F_{\gamma}$ with a flatter SNR-produced CRP distribution ($\alpha_{\rm SN}=4.3$), but a steep SW-produced CRP distribution ($\alpha_{\rm SW}=4.5$). Here, $p_{\rm max}=100$ PeV/$c$ is used in both the SNR and SW-produced CRP distributions (also in the following Figures \ref{fig:f5} to \ref{fig:f7}); the value of $p_{\rm max}$ is not important in the $\gamma$-ray energy range shown, due to the attenuation by extragalactic background photos, and the results remain the same for $p_{\rm max}=1~{\rm PeV}/c$. Observational data (dots with error bars) are shown for comparison. For M82, the Fermi-LAT data are taken from \citet{acero2015} and the Veritas data come from \citet{veritas2009}. For NGC253, both Fermi-LAT and H.E.S.S. data are taken from \citet{hess2018}. For Arp220, the Fermi-LAT data are taken from \citet{peng2016} and the upper limit by Veritas comes from \citet{fleischhack2015}. (The same observational data are shown in the following Figures \ref{fig:f5} to \ref{fig:f7}.) \label{fig:f4}}
\end{figure*}

\begin{figure*}[t]
\vskip -0.4 cm
\hskip -0.4 cm
\centerline{\includegraphics[width=1.05\textwidth]{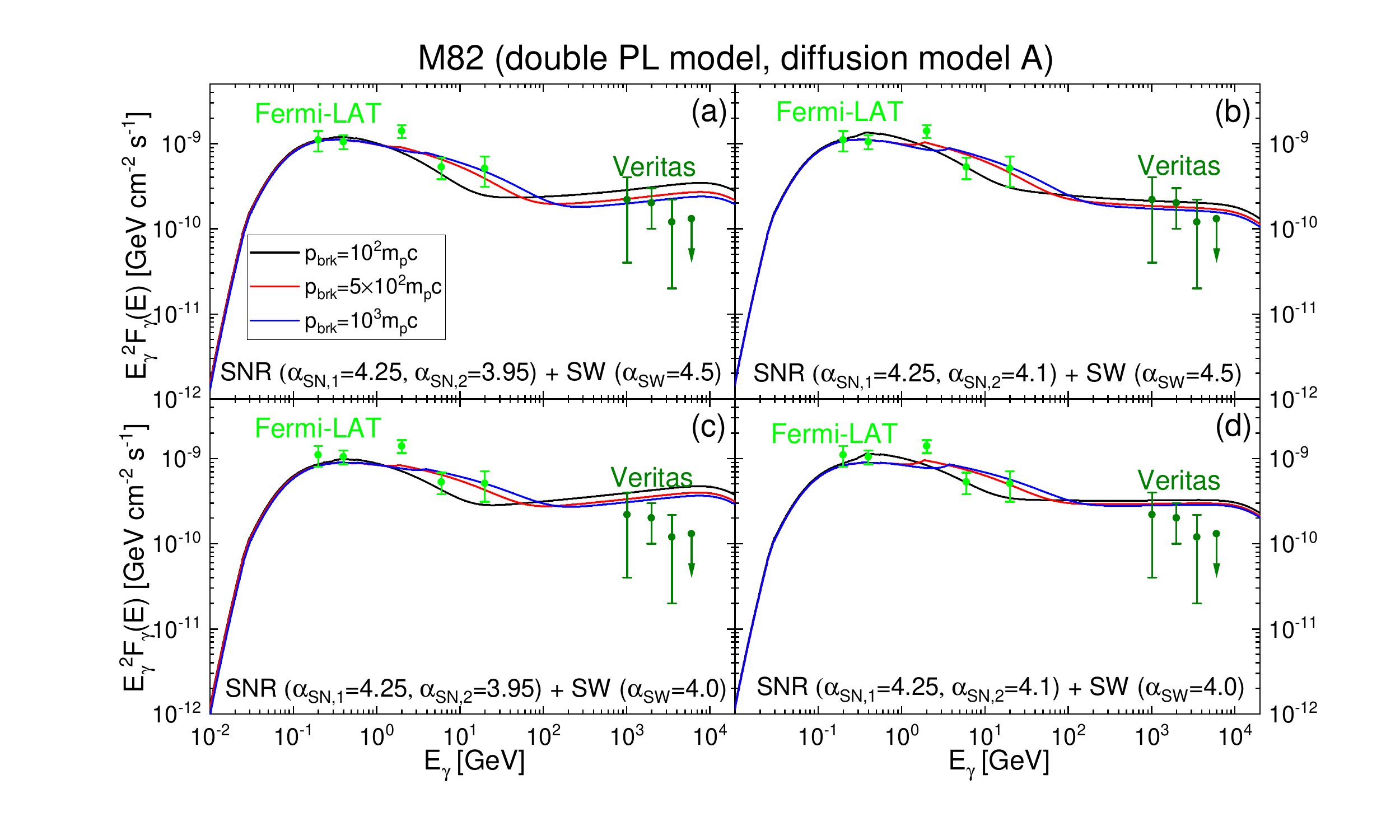}}
\vskip -1. cm
\caption{Comparison of estimated $\gamma$-ray fluxes with observational data (dots with error bars) for M82. The SNR-produced CRPs are modeled with double PL momentum distributions, while the SW-produced CRPs are modeled with single PL momentum distributions. 
Here, $p_{\rm max}=100~{\rm PeV}/c$ for both the SNR and SW-produced CRP distributions, and different values of $p_{\rm brk}$ are tried for the SNR-produced CRP distributions.
For turbulent diffusion, the model A is used.
\label{fig:f5}}
\end{figure*}

\begin{figure*}[t]
\vskip -0.4 cm
\hskip -0.4 cm
\centerline{\includegraphics[width=1.05\textwidth]{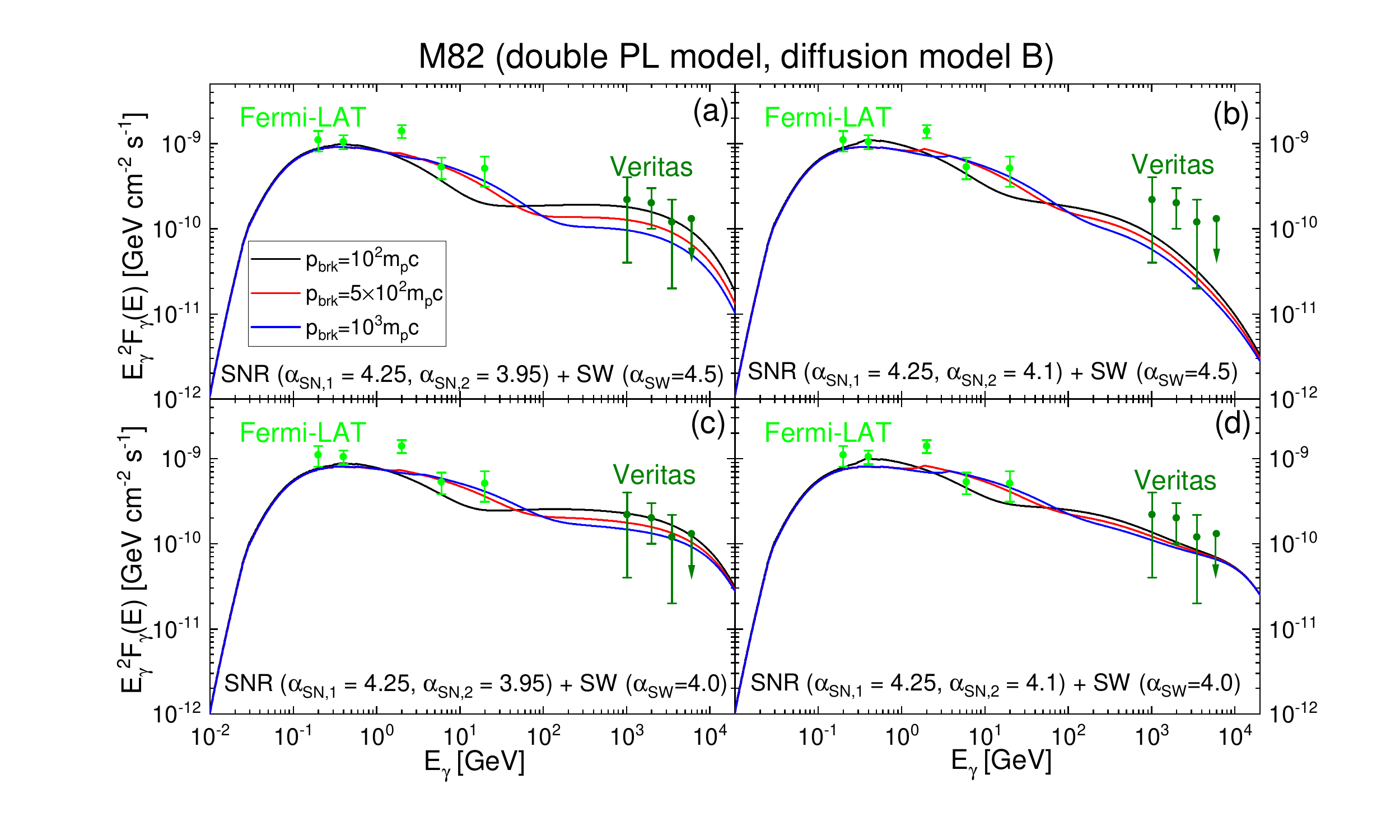}}
\vskip -1. cm
\caption{Same as in Fig. 5, except using the diffusion model B.
\label{fig:f6}}
\end{figure*}

\subsection{Combined Cosmic-Ray Productions}
\label{sec:s2.5}

Using $\mathcal{N}_{\rm SN,p}(p)$ and $\mathcal{N}_{\rm SW,p}(p)$, we define the CRP injection rate in the SBN as 
\begin{equation}
Q_p(p) =  \frac{\mathcal{N}_{\rm SN,p}(p) + \mathcal{N}_{\rm SW,p}(p)}{V},
\label{eq8}
\end{equation}
where $V$ is the volume of the SBN, calculated with $R_{\rm SBN}$ in Table \ref{tab:t1}. We have two CRP production models, (1) the single PL model, in which both the SNR and SW-produced CRPs follow single PL distributions, 
and (2) the double PL model, in which the SNR-produced CRPs follow double PL distributions while the SW-produced CRPs follow single PL distributions. 

Figure \ref{fig:f2} compares $Q_p(p)$ of three nearby SBGs for the two CRP production models (single versus double PLs) 
with two different slopes of SW-produced CRPs, $\alpha_{\rm sw}=4.0$ (blue lines) and 4.5 (red lines), 
covering the range in the two opposite limits.
The consequence of the double PL model is obvious, especially in the high-momentum range. For M82, for instance, $Q_p(p)$ of the double PL model (panel (d)) is several to ten times larger than that of the single PL model (panel (a)) for $p \gtrsim 0.1$ PeV/$c$. This implies that the flux of high energy neutrinos of $E_{\nu} \gtrsim 10$ TeV would be larger with the double PL model (see Section \ref{sec:s5}). 
With a flat spectrum, such as $\alpha_{\rm SW} \simeq 4.0$, the SW-produced CRPs could substantially contribute to $Q_p(p)$ over a wide momentum range. Especially, in Arp220 where the cases of $\beta = 1$ are shown, the SFR is high and the SW contribution could be quite important, especially at high momenta. On the other hand, with a much steeper spectrum, such as $\alpha_{\rm SW} \simeq 4.5$, the SW contribution should be less noticeable. 

\subsection{Comments on the Cosmic-Ray Production Model}
\label{sec:s2.6}

In this subsection, we further address the justification of our CRP production model, in particular, the double PL form for SNR-produced CRPs. As described in the introduction, in numerical studies of nonlinear DSA, the presence of a concavity in the cumulative CRP spectrum integrated during the SNR evolution, and hence the double PL spectrum of CRPs, are still under controversy. \citet{vladimirov2008}, \citet{zirakashvili2012}, and \citet{kang2013}, for instance,  showed that the cumulative CRP spectrum may retain only a mild concave curvature in their numerical simulations, depending on the Alfv\'enic drift model. However, in a recent work, \citet{caprioli2020} suggested that the CRP spectrum could become softer than the canonical DSA spectrum, and hence the concavity might weaken if the downstream Alfv\'enic drift is to be efficient. The concave curvature in the cumulative CRP spectrum seems to depend on the details of the phenomenological models employed in numerical studies, and hence its presence has yet to be confirmed.

We note that there is no observational evidence for a concave structure in the CRP spectrum of our Galaxy. The diffuse $\gamma$-ray spectrum from the Galactic disk, for instance, shows a smooth structure, which would be consistent with the CRP spectrum without a concavity \citep[e.g.,][]{yang2016}. However, the double PL flux model has been commonly employed to fit the observed $\gamma$-ray fluxes of SBGs. In the case of NGC253, for instance, the slope of the H.E.S.S.-observed $\gamma$-ray spectrum looks different from that obtained with Fermi-LAT \citep[e.g.,][]{abramowski2012, hess2018}. (See also, e.g., \citet{tamborra2014} and \citet{bechtol2017}, and the discussion in the introduction.) Unlike in SBGs, in our Galaxy a number of additional processes would be possibly involved. For instance, the $\gamma$-ray spectrum of the Galactic disk may steepen due to the rigidity-dependent escape of CRPs. In addition, the superposition of CRPs of different origins may make the $\gamma$-ray spectrum smooth. 
In SBGs, the efficient CRP production at SNRs owing to high SN explosion rates could lead to the cumulative CRP spectrum that is different from that of our Galaxy.

Here, in addition to the canonical single PL model, we also consider the double PL model for SNR-produced CRPs. We suggest that the observed $\gamma$-ray fluxes of nearby SBGs including the break could be reproduced with both the models in the combination of different transport models presented in next section.

\section{Models for Cosmic-Ray Transport}
\label{sec:s3}

Following the simple approach adopted in previous studies (references in the introduction), we assume that the production of CRPs is balanced by the energy loss, SBN wind advection, and diffusion. Then, the momentum spectrum of the CRP density in the SBN, $f_p(p)$, is given as
\begin{equation}
\frac{f_p(p)}{\tau_{\rm loss}}+\frac{f_p(p)}{\tau_{\rm adv}}+\frac{f_p(p)}{\tau_{\rm diff}} = Q_p(p).
\label{eq9}
\end{equation}
The energy loss is mostly due to ionization, Coulomb collisions, and inelastic $pp$ collisions, and for its time-scale $\tau_{\rm loss}$, we use the formulae in \citet{peretti2019}. The wind advection time-scale, $\tau_{\rm adv}$, is calculated as the radius of the SBN divided by the escaping speed of the SBN wind, i.e., $R_{\rm SBN}/v_{\rm SBNwind}$ with $R_{\rm SBN}$ and $v_{\rm SBNwind}$ given in Table \ref{tab:t1}.
Figure \ref{fig:f3} shows $\tau_{\rm loss}$ (magenta solid line) and $\tau_{\rm adv}$ (black dashed line)
estimated for M82; the two time-scales are comparable.

For the diffusion time-scale $\tau_{\rm diff}$, we consider two different diffusion models. First, we adopt the model A of \citet{peretti2019}, where CRPs scatter off the large-scale Kolmogorov turbulence in the SBN (hereafter, the diffusion model A). Then, the diffusion coefficient is given as 
\begin{equation}
D_{\rm A}(p) \approx \frac{r_g(p) c}{3\mathcal{F}(k)},
\label{eq10}
\end{equation}
where $r_g(p)$ is the gyroradius of CRPs and the speed of CRPs is assumed to be $c$. $\mathcal{F}(k)$ is the normalized energy density of turbulent magnetic fields per unit logarithmic wavenumber, which is defined as
\begin{equation}
\int_{k_0}^\infty \mathcal{F}(k) \frac{dk}{k} = \left(\frac{\delta B_{\rm SBN}}{B_{\rm SBN}}\right)^2 =1
\label{eq11}
\end{equation}
with $k_0=1~{\rm pc}^{-1}$. Note that 1 pc is about the gyroradius of CRP of 100 PeV in the background magnetic field of 100 $\mu$G. The diffusion time-scale is given as $\tau_{\rm diff} = R_{\rm SBN}^2/D_{\rm A}(p)$ with the radius of the SBN. We point that $r_g(p) \propto p$, and $\mathcal{F}(k) \propto k^{-2/3} \propto p^{2/3}$ assuming that CRPs interact with the resonant mode corresponding to the gyroradius, and then $\tau_{\rm diff} \propto p^{-1/3}$. \citet{peretti2019} also considered the Bohm diffusion, for which $D(p) \sim r_g(p)c \propto p$ and $\tau_{\rm diff} \propto p^{-1}$. But they argue that the Bohm diffusion essentially gives the same results as the Kolmogorov diffusion for the transport of CRPs, because the SBN wind advection has a shorter time-scale over most of the momentum range (see Figure \ref{fig:f3}) and so those diffusions are subdominant. Hence, we here consider only the Kolmogorov turbulence for the diffusion model A.

We also consider the diffusion model of \citet{krumholz2020}, where the CRPs of $E_{\rm CR} \lesssim 0.1$ PeV resonantly scatter off self-excited turbulence, assuming that the ISM turbulence of scales less than the gyroradius of CRPs with $\lesssim 0.1~{\rm PeV}$ is wiped out by ion-neutral damping (hereafter, the diffusion model B). Assuming that the turbulence is trans or super-Alfv\'enic, the diffusion coefficient is given as
\begin{equation}
D_{\rm B}(p) \approx v_{\rm st}\ell ,
\label{eq12}
\end{equation}
where $v_{\rm st}$ is the CRP streaming speed, $\ell \approx \ell_0/M_{\rm A, turb}^3$, $\ell_{0}$ is the outer scale of the turbulence, and $M_{\rm A,turb}$ is the Alfv\'en Mach number of the turbulence. 
Following \citet{krumholz2020}, $\ell_{0}=H_{\rm gas}$, i.e., the gas scale height listed in Table \ref{tab:t1}, is used, and $M_{\rm A,turb}=2$ is assumed. 
And $v_{\rm st}$ is modeled as ($v_{\rm st}/v_{\rm Ai}-1) \propto {(E_{\rm CR}/{\rm GeV})}^{\alpha - 3}$, where $v_{\rm Ai}$ is the ion Alfv\'en speed given in Table \ref{tab:t1} and $\alpha$ is the PL slope of CRP momentum distribution, $\alpha_{\rm SN}$ or $\alpha_{\rm SW}$, described in Section \ref{sec:s2}.
The proportional constant for different SBGs can be found in Table 1 of \citet{krumholz2020}. 
The diffusion time-scale is given as $\tau_{\rm diff} = H_{\rm gas}^2/D_{\rm B}(p)=M_{\rm A, turb}^3 H_{\rm gas} / v_{\rm st}$.

Figure \ref{fig:f3} compares $\tau_{\rm diff}$ to $\tau_{\rm loss}$ and $\tau_{\rm adv}$ for M82. 
In the estimation of $\tau_{\rm diff}$ of the diffusion model B in the figure (blue solid line), the single PL momentum distribution of SNR-produced CRPs only with $\alpha_{\rm SN}=4.25$ is used. 
The slope of $\tau_{\rm diff}(p)$ is $(3-\alpha_{\rm SN})$ in the range of $p\sim1-100$ TeV/$c$ for the diffusion model B. 
With the diffusion model A, the energy loss and wind advection dominate over the turbulent transport, except at the highest momentum of $p \gtrsim 10$ PeV/$c$. 
In the case of the diffusion model B, on the other hand, $\tau_{\rm diff}$ becomes shorter than $\tau_{\rm loss}$ 
or $\tau_{\rm adv}$ for $p \gtrsim 1$ TeV/$c$, so high energy CRPs are confined within the SBN for shorter time and 
have less chance for collisions with background thermal protons. 
The kink in $\tau_{\rm diff}$ of the diffusion model B appears at $p \sim 0.1$ PeV/$c$, because the momentum-dependent streaming velocity becomes greater than the speed of light.
Hence, the model stops being valid at $p \gtrsim 0.1$ PeV/$c$; the CRPs in this high momentum range may interact with the large-scale turbulence in the SBN, resulting in longer $\tau_{\rm diff}$. This may affect in particular the estimation of neutrino emission in $E_{\nu} \gtrsim 10$ TeV (see Section \ref{sec:s5} for further comments).

\begin{figure}[t]
\vskip -0.5 cm
\hskip 0.2 cm
\centerline{\includegraphics[width=0.55\textwidth]{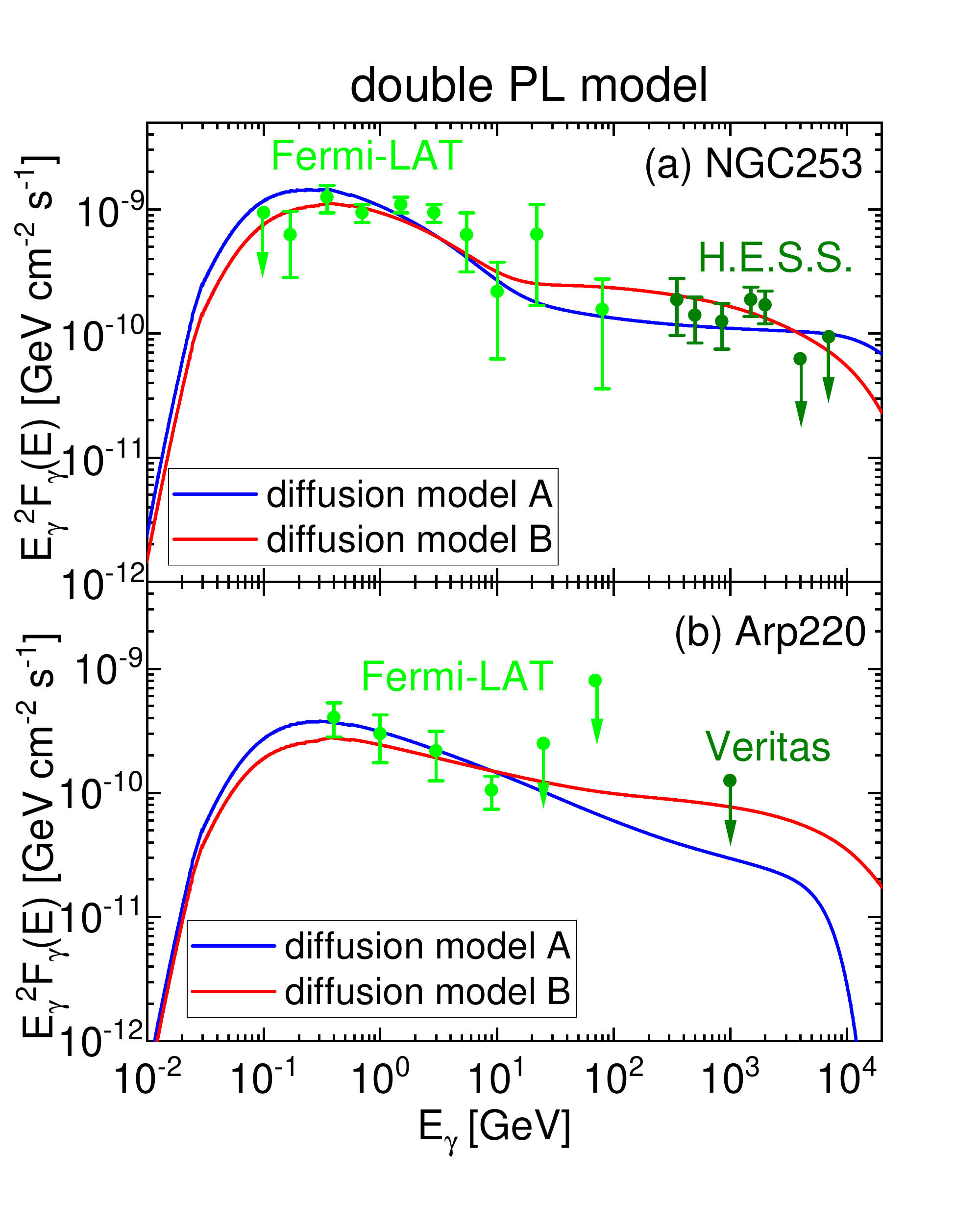}}
\vskip -0.8 cm
\caption{Comparison of estimated $\gamma$-ray fluxes with observational data (dots with error bars) for NGC253 in panel (a) and Arp220 in panel (b), using two different diffusion models. The SNR-produced CRPs are modeled with double PL momentum distributions, while the SW-produced CRPs are modeled with single PL momentum distributions, with the following slopes: 
for NGC253, $\alpha_{\rm SN,1}=4.3$, $\alpha_{\rm SN,2}=4.1$, $\alpha_{\rm SW}=4.5$ (model A, blue line), and $\alpha_{\rm SN,1}=4.3$, $\alpha_{\rm SN,2}=3.95$, $\alpha_{\rm SW}=4.0$ (model B, red line); 
for Arp220, $\alpha_{\rm SN,1}=4.45$, $\alpha_{\rm SN,2}=4.1$, $\alpha_{\rm SW}=4.5$ (model A, blue line), and $\alpha_{\rm SN,1}=4.45$, $\alpha_{\rm SN,2}=3.95$, $\alpha_{\rm SW}=4.0$ (model B, red line).
For Arp220, the cases of $\beta = 1$ are shown.
Here, $p_{\rm max}=100~{\rm PeV}/c$ and $p_{\rm brk} = 10^2~m_p c$ for both the SNR and SW-produced CRP distributions.
\label{fig:f7}}
\end{figure}

\section{Gamma-ray Emission from SBGs}
\label{sec:s4}

With the CRP momentum distribution function, $f_p(p)$, calculated in Equation (\ref{eq9}), we first estimate the $\gamma$-ray emission due to $pp$ collisions.
As in previous studies, we use the pion source function as a function of pion energy $E_{\pi}$, presented in \citet{kelner2006},
\begin{equation}
q_{\pi}(E_{\pi})=\frac{cn_{\rm SBN}}{K_{\pi}}\sigma_{\rm pp}\left(m_p c^2 + \frac{E_{\pi}}{K_{\pi}}\right)n_p\left(m_p c^2 + \frac{E_{\pi}}{K_{\pi}}\right).
\label{eq13}
\end{equation}
Here, $K_{\pi} \sim 0.17$ is the fraction of the kinetic energy transferred from proton to pion, and $n_p(E)$ is the CRP energy distribution, given as $n_p(E)dE = 4\pi p^2 f_p(p)dp$. And $\sigma_{\rm pp}(E)$ is the cross-section, given as $(34.3 + 1.88 \theta + 0.25 \theta^2)\times[1-(E_{\rm th}/E)^4]^2$ mb, where $\theta = \ln(E/{\rm TeV})$ and $E_{\rm th}=1.22$ GeV is the threshold energy. The $\gamma$-ray source function as a function of $\gamma$-ray energy $E_{\gamma}$ is calculated as
\begin{equation}
q_{\gamma}(E_{\gamma})=2\int_{E_{\rm min}}^{\infty}\frac{q_{\pi}(E_{\pi})}{\sqrt{E_{\pi}^2 - m_{\pi}^2 c^4}}dE_{\pi},
\label{eq14}
\end{equation}
where $E_{\rm min}=E_{\gamma}+m_{\pi}^2 c^4/(4E_{\gamma})$. Then, the $\gamma$-ray flux as a function of $E_{\gamma}$ is given as
\begin{equation}
F_{\gamma}(E_{\gamma})=\frac{q_{\gamma}(E_{\gamma})V}{4\pi D_{\rm L}^2}\exp\left[-\tau_{\gamma\gamma}(E_{\gamma},z)\right],
\label{eq15}
\end{equation}
where $D_{\rm L}$ is the luminosity distance of SBGs in Table \ref{tab:t1}. 
Here, the optical depth, $\tau_{\gamma\gamma}$, takes into account pair production with the background photons during the propagation in the intergalactic space, and we use the analytic approximation described in the Appendix C. of \citet{peretti2019} \citep[see also][]{franceschini2017}.

Figure \ref{fig:f4} shows $F_{\gamma}$, estimated for nearby SBGs with the single PL distributions for both SNR-produced CRPs and SW-produced CRPs. The best fitted values of \citet{peretti2019} are chosen for $\alpha_{\rm SN}$, while different values are tried for $\alpha_{\rm SW}$. The upper panels (a) - (c) show $F_{\gamma}$ with the diffusion model A. The $\gamma$-ray observations are well reproduced. The contribution of SW CRPs is subdominant for M82 and NGC253, as expected with $\mathcal{L}_{\rm SW} \sim 0.3\mathcal{L}_{\rm SN}$ in Table \ref{tab:t1}. However, with larger $\mathcal{L}_{\rm SW} \sim (0.43 - 0.87)\mathcal{L}_{\rm SN}$ for Arp220, we argue that the contribution of SW CRPs could be important, especially if $\alpha_{\rm SW}$ is small;
for instance, with $\alpha_{\rm SW}=4.0$ and $\beta=1$, $F_{\gamma}$ at $E_{\gamma} \sim 1$ TeV approaches the upper limit of the Veritas observation.
The bottom panels (d) - (f) show $F_{\gamma}$ with the diffusion model B. Our estimation of $F_{\gamma}$ is consistent with \citet{krumholz2020} (see their figure 4), although the CRP advection by the SBN wind is additionally included in our CRP transport model (\citet{krumholz2020} did not include the wind advection). The Fermi-LAT $\gamma$-ray observations, especially those of M82 and NGC253, are reproduced regardless of the diffusion model. However, $F_{\gamma}$ at $E_{\gamma} \gtrsim 1$ TeV is sensitive to diffusion model. Overall, while $F_{\gamma}$ estimated with the diffusion model A in the upper panels looks more consistent with observations, the case with the diffusion model B in the bottom panels may not be ruled out with current observations (see below).

We note that while $F_{\gamma}$ with $\alpha_{\rm SN} = 4.2 - 4.3$ fits the Fermi-LAT $\gamma$-ray fluxes of M82 and NGC253, those values would be too small to explain the steeper spectrum of Arp220 \citep[e.g.,][]{peretti2019}. If a steep SW contribution of $\alpha_{\rm SW} = 4.5$ is included, the Fermi-LAT observation of Arp220 may be reproduced even with $\alpha_{\rm SN} = 4.3$. 
The gray solid lines in panels (c) and (f) of Figure \ref{fig:f4} show the predicted flux $F_{\gamma}$ with $\alpha_{\rm SN} = 4.3$ and $\alpha_{\rm SW} = 4.5$. But then, the flux at $E_{\gamma} \gtrsim 1$ TeV seems to be much smaller than the Veritas upper limit, by an order of magnitude or more.

\begin{deluxetable}{lccccccccccccccccccc}[t]
\tablecaption{Results of $\chi^2$-Test with the Best Fit Model Parameters for Combinations of Different CRP Production and Diffusion Models \label{tab:t2}}
\tablenum{2}
\tablehead{
\colhead{~} &
\colhead{~} &
\colhead{SNR CRPs}&
\colhead{~} &
\colhead{diffusion}&
\colhead{$\chi^2$} &
\colhead{$p$-value$^a$}}
\startdata
M82    && single PL  && A  & 4.96  & 0.54\\
       && single PL  && B  & 8.95  & 0.17\\
       && double PL  && A  & 6.11 & 0.19\\
       && double PL  && B  & 5.95  & 0.21\\
\hline
NGC253 && single PL  && A  & 11.51 & 0.48\\
       && single PL  && B  & 17.04 & 0.15\\
       && double PL  && A  & 11.34  & 0.33\\
       && double PL  && B  & 9.68  & 0.47\\
\enddata
\tablenotetext{a}{The $p$-value is defined as the probability of having the $\chi^2$-value larger than the estimated $\chi^2$-value.}
\end{deluxetable}

The $\gamma$-ray fluxes of M82 estimated with the double PL momentum distributions for SNR CRPs are shown in Figure \ref{fig:f5} (with the diffusion model A) and Figure \ref{fig:f6} (with the diffusion model B).
The single PL momentum distributions with $\alpha_{\rm SW}=4.0$ or 4.5 are employed for SW CRPs. 
The predicted flux $F_{\gamma}$ in the energy range of $E_{\gamma} \lesssim 100$ GeV is insensitive to the diffusion model, since the transport of CRPs with $p \lesssim 1$ TeV/$c$ is controlled by the SBN wind advection. 
Hence, the Fermi-LAT data are reproduced for both of the diffusion models, if the slope of the low momentum part is adjusted, i.e., $\alpha_{\rm SN,1} = 4.25$. 
On the other hand, $F_{\gamma}$ in the higher energy range is affected substantially by the diffusion model,
because the transport time-scale of CRPs with $p \gtrsim 1$ TeV/$c$ is much longer in the diffusion model A than in the diffusion model B, as shown in Figure \ref{fig:f3}. 
The reproduction of the Veritas data for $E_{\gamma} \gtrsim 1$ TeV requires a softer spectrum in the high momentum part, e.g., $\alpha_{\rm SN,2} = 4.1$ (see Fig. \ref{fig:f5}(b)), in the case of the diffusion model A, while it requires a harder spectrum, e.g., $\alpha_{\rm SN,2} = 3.95$ (see Fig. \ref{fig:f6}(c)), in the case of the diffusion model B. Other model parameters involved are less important. 
Yet, somewhat better fittings are obtained with larger $\alpha_{\rm SW}$, such as 4.5, if the diffusion model A is adopted, and with smaller $\alpha_{\rm SW}$, such as 4.0, if the diffusion model B is adopted.
The break momentum $p_{\rm brk}$ affects the predicted fluxes at high energies, but with the adopted range of $p_{\rm brk} \simeq 10^2 - 10^3 m_p c$, the difference in $F_{\gamma}$ is at most a factor of two.

Figure \ref{fig:f7} shows the $\gamma$-ray fluxes of NGC253 and Arp220 estimated with the double PL distributions for SNR CRPs. 
For $\alpha_{\rm SN,1}$, the values that reproduce the Fermi-LAT data are used.
Again, with the diffusion model A (blue lines), a softer distribution at the high momentum part with $\alpha_{\rm SN,2}=4.1$ reproduces the the H.E.S.S. and Veritas observations better, while with the diffusion model B (red lines), a harder distribution with $\alpha_{\rm SN,2}=3.95$ does better. The model parameters used are listed in the figure caption.

In general, at $E_{\gamma} \lesssim 100$ GeV, the $\gamma$-ray observations by Fermi-LAT are well reproduced, regardless of the CRP production and diffusion models, once the slope of the SNR-produced CRP distribution, $\alpha_{\rm SN}$ or $\alpha_{\rm SN,1}$, is properly chosen. At $E_{\gamma} \gtrsim 1$ TeV, on the other hand, the results depend rather sensitively on both the CRP production and diffusion models. To examine the goodness of the fit to the $\gamma$-ray observations by Fermi-LAT, Veritas and H.E.S.S., we perform the $\chi^2$-test for M82 and NGC253 with enough data points, by calculating $\chi^2 \equiv \sum [(F_{\rm \gamma,obs} - F_{\gamma})/F_{\rm \gamma,err}]^2$, where $F_{\gamma}$ is our estimated $\gamma$-ray flux, $F_{\rm \gamma,obs}$ and $F_{\rm \gamma,err}$ are the observed $\gamma$-ray flux and error, respectively; only the data points with error bar are used, while those of upper limit are excluded. The $\chi^2$-values and $p$-values for different combinations of the CRP production and diffusion models with the best fit parameters are given in Table \ref{tab:t2}. For both M82 and NGC253, the combination of the single PL model and the diffusion model A and the combination of the double PL model and the diffusion model B give better statistics. However, other combinations also have, for instance, the $p$-value within the 1 $\sigma$ level, and hence should not be statistically excluded. On the other hand, the number of observation data points, especially that for M82, seems to too small to differentiate different models. Our results imply that further $\gamma$-ray observations would be necessary to meaningfully constrain the CRP production and transport models in SBGs.

\begin{deluxetable}{lcccccccccccccc}[t]
\tablecaption{Best Fit Model Parameters for Reproduction of $\gamma$-ray Observations \label{tab:t3}}
\tablenum{3}
\tablehead{
\colhead{~} &
\colhead{SNR CRPs}&
\colhead{$\alpha_{\rm SN,1}$}&
\colhead{$\alpha_{\rm SN,2}$}&
\colhead{$\alpha_{\rm SW}$}&
\colhead{diffusion}}
\startdata
M82&    single PL & 4.25   & -     & 4.5 & A\\
NGC253& single PL & 4.3    & -     & 4.5 & A\\
Arp220& single PL & 4.45   & -      & 4.5 & A\\
\hline
M82&    double PL & 4.25 & 4.1   & 4.5 & A\\
NGC253& double PL & 4.3 & 4.1   & 4.5 & A\\
Arp220& double PL & 4.45 & 4.1 & 4.5 & A\\
\hline
M82&    double PL & 4.25 & 3.95 & 4.0 & B\\
NGC253& double PL & 4.3 & 3.95   & 4.0 & B\\
Arp220& double PL & 4.45 & 3.95 & 4.0 & B\\
\enddata
\tablenotetext{}{$p_{\rm brk}=100~m_{\rm p}c$ is used for the double PL distribution. For Arp220, $\beta = 1$ is used.}
\end{deluxetable}

\begin{figure*}[t]
\vskip -0.7 cm
\hskip -0.8 cm
\centerline{\includegraphics[width=1.1\textwidth]{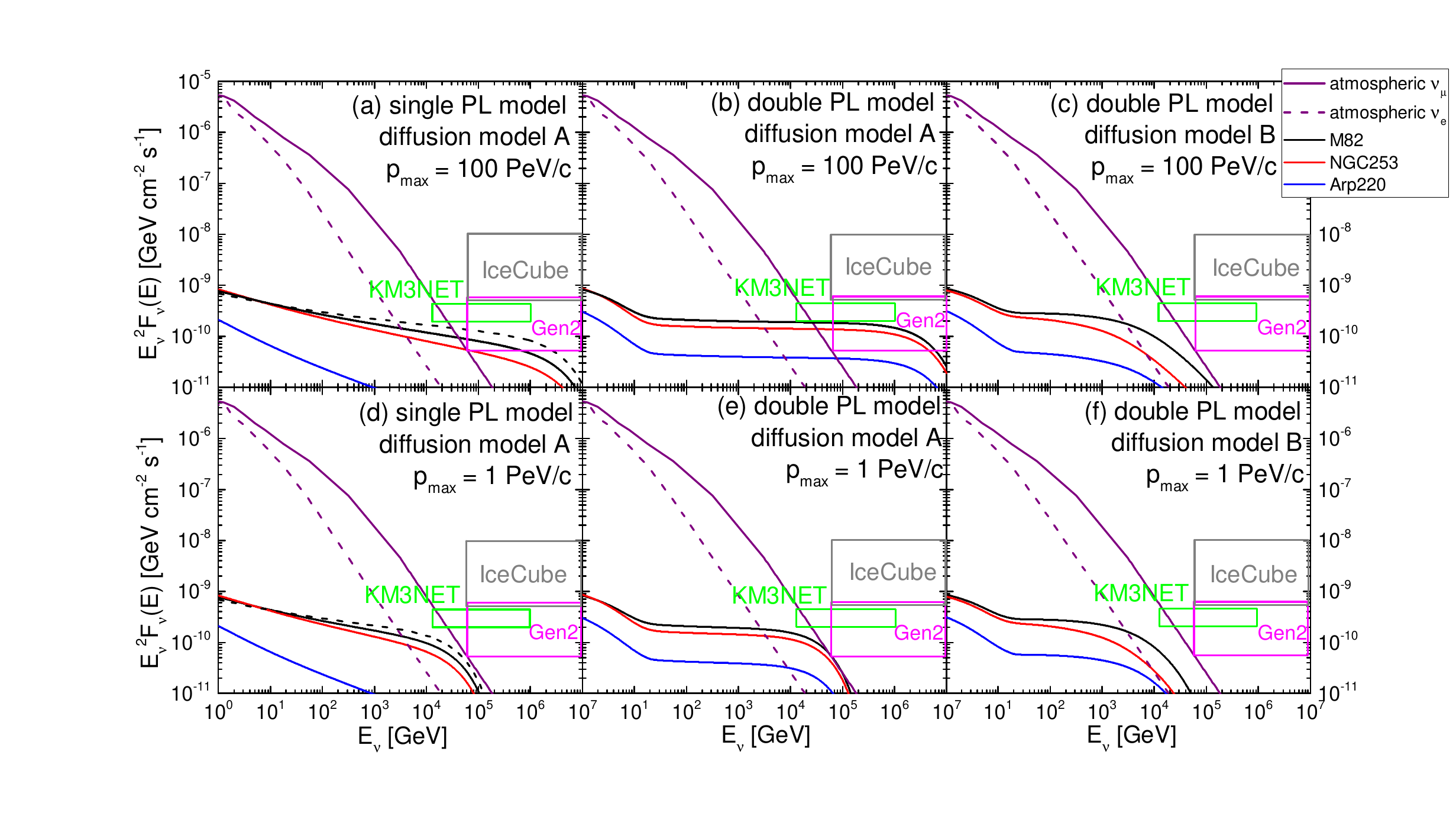}}
\vskip -1. cm
\caption{Predicted neutrino fluxes from three nearby SBGs with the model parameters listed in Table \ref{tab:t3} (solid lines); $p_{\rm max}=100~{\rm PeV}/c$ (upper panels) and $p_{\rm max}=1~{\rm PeV}/c$ (lower panels) are used.
In panels (a) and (d), the black solid line is for M82 with $\alpha_{\rm SN}=4.25$, while the black dashed line is with $\alpha_{\rm SN}=4.2$.
The boxes draw the ranges of the IceCube point source sensitivity \citep[][grey]{aartsen2017}, the IceCube-Gen2 point source sensitivity \citep[][magenta]{santen2017} the KM3NET point source sensitivity \citep[][green]{aiello2019} (see the text for details). The dark purple solid and dashed lines draw the fluxes of atmospheric muon and electron neutrinos from the data of \citet{richard2016}. The fluxes shown are within a circular beam of $1^{\circ}$ diameter. \label{fig:f8}}
\end{figure*}

\section{Neutrino Emission from SBGs}
\label{sec:s5}

We then predict the neutrino fluxes from three nearby SBGs for three different combinations of CRP production and transport models with the best fit model parameters listed in Table \ref{tab:t3}. The combination of the single PL model for SNR CRPs and the diffusion model B is not included, since it shows the largest deviation from observation data. We again employ the formulae presented in \citet{kelner2006}. The neutrino source function as a function of neutrino energy $E_{\nu}$ is given as
\begin{equation}
q_{\nu}(E_{\nu})=2\int_0^1\left[f_{\nu_{\mu}^{(1)}}(x)+f_{\nu_{\mu}^{(2)}}(x)+f_{\nu_e}(x)\right]q_{\pi}\left(\frac{E_{\nu}}{x}\right)\frac{dx}{x},
\label{eq16}
\end{equation}
where $x=E_{\nu}/E_{\pi}$. The term $f_{\nu_{\mu}^{(1)}}(x)$ represents the muon neutrino production through the pion decay $\pi \rightarrow \mu \nu_{\mu}$, and $f_{\nu_{\mu}^{(2)}}(x)$ and $f_{\nu_{e}}(x)$ describe the muon and electron neutrino productions through the muon decay $\mu \rightarrow \nu_{\mu} \nu_{e} e$. The formulae for $f_{\nu_{\mu}^{(1)}}$, $f_{\nu_{\mu}^{(2)}}$, and $f_{\nu_{e}}$ are given in \citet{kelner2006}. 
Then, the neutrino flux observed at the Earth is calculated as
\begin{equation}
F_{\nu}(E_{\nu})=\frac{q_{\nu}(E_{\nu})V}{4\pi D_{\rm L}^2}.
\label{eq17}
\end{equation}

Figure \ref{fig:f8} shows the predicted neutrino fluxes, $F_{\nu}$, for two values of the maximum momentum of CRP distributions, 
$p_{\rm max}=100~{\rm PeV}/c$ (upper panels) and $p_{\rm max}=1~{\rm PeV}/c$ (lower panels).
We compare $F_{\nu}$ to the point source sensitivity ranges of IceCube, IceCube-Gen2, and KM3NET: $E_{\nu}^2F_{\nu} \sim 5 \times 10^{-10} - 10^{-8}~{\rm GeV~cm^{-2}s^{-1}}$ in the energy range of $E_{\nu} \gtrsim 60$ TeV for IceCube with seven-year data \citep[][]{aartsen2017}, $E_{\nu}^2F_{\nu} \sim 5 \times (10^{-11} - 10^{-10})~{\rm GeV~cm^{-2}s^{-1}}$ in the energy range of $E_{\nu} \gtrsim 60$ TeV for IceCube-Gen2 with fifthteen-year data \citep[][]{santen2017}, and $E_{\nu}^2F_{\nu} \sim (2 - 4) \times 10^{-10}~{\rm GeV~cm^{-2}s^{-1}}$ in the energy range of $E_{\nu} \sim 15$ TeV $-$ 1 PeV expected for KM3NET with six-year data \citep[][]{aiello2019}. The point source sensitivity in the figure is defined as the median upper limit at the 90\% confidence level. We also compare $F_{\nu}$ to the atmospheric muon and electron neutrino fluxes \citep[][]{richard2016}. A circular beam of $1^{\circ}$ diameter is used for the calculation of the fluxes shown in the figure.

With the adopted range of $p_{\rm max} \simeq 1-100$ PeV/$c$, the $\gamma$-ray fluxes presented in the previous section are insensitive to $p_{\rm max}$, due to the exponential attenuation included in Equation (\ref{eq15}).
However, $F_{\nu}(E_{\nu})$ for high energy neutrinos exhibits strong dependence on the assumed value of $p_{\rm max}$.
With larger $p_{\rm max}$, $F_{\nu}$ extends to higher energies. But the amount of highest energy neutrinos is determined not just by $p_{\rm max}$, but also by the diffusion model. If the diffusion time-scale of CRPs is shorter at high energies, $F_{\nu}$ should be lower. 
Hence, $F_{\nu}$ is smaller at highest energies with the diffusion model B than with the diffusion model A in Figure \ref{fig:f8}. As mentioned in Section \ref{sec:s3}, in the diffusion model B, the streaming speed of CRPs approaches the speed of light at $p \gtrsim 0.1$ PeV/$c$; that is, the streaming speed may be overestimated and the diffusion time-scale may be underestimated, and then $F_{\nu}$ may be underestimated at highest energies. 
Yet, in all the models considered in Figure \ref{fig:f8}, which fit $\gamma$-ray observations reasonably well, the predicted $F_{\nu}$ at $E_{\nu} \gtrsim 0.1$ PeV from nearby SBGs is at least an order of magnitude smaller than the lower bound of the point source sensitivity of IceCube. 
Thus, SBGs may not be observed as point sources in IceCube. SBGs could be still a major contributor of diffuse high-energy neutrinos, as pointed in previous works (see the introduction for references), but the amount of estimated contributions should depend on the details of modeling including the diffusion model.

Our results suggest that nearby SBGs, especially M82 and also NGC253, might be observed as point sources in the range of 15 TeV $\lesssim E_{\nu} \lesssim 1$ PeV with KM3NET and also in the range of 60 TeV $\lesssim E_{\nu} \lesssim 10$ PeV with IceCube-Gen2, in the most optimistic cases with the diffusion model A and $p_{\rm max}=100$ PeV/$c$, shown in panels (a) and (b) of Figure \ref{fig:f8}. On the other hand, other models predict $F_{\nu}$ that does not touch the point source sensitivities boxes of KM3NET and IceCube-Gen2, indicating that nearby SBGs would be difficult to be detected as point sources. We point that the observations of neutrinos from nearby SBGs would help constrain the CRP production and transport models there.

At lower energies of $E_{\nu}\lesssim1$ TeV, the neutrino flux has been obtained, for instance, in Super-Kamiokande \citep[e.g.,][]{hagiwara2019}. However, $F_{\nu}$ from nearby SBGs is predicted to be orders of magnitude smaller than the atmospheric muon neutrino flux. So it would not be easy to separate the signature of neutrinos from SBGs in the data of ground facilities such as Super-Kamiokande and future Hyper-Kamiokande \citep[e.g.,][]{abe2011}.

\section{Summary}
\label{sec:s6}

We have estimated the CRP production by SWs from massive stars and 
SNRs from core-collapse SNe inside the SBN of nearby SBGs. Considering the lack of full understanding of nonlinear interrelationships among complex kinetic processes involved in
the DSA theory, we have adopted both the canonical single PL model with $\alpha_{\rm SN}$ and 
the double PL model with $\alpha_{\rm SN,1}$ and $\alpha_{\rm SN,2}$ for the SNR-produced CRP momentum distribution.
The latter is intended to represent a concave spectrum that might arise due to nonlinear dynamical feedback from the CR pressure,
if $\sim 10$ \% of the shock kinetic energy is transferred to CRPs.
For the SW-produced CRP distribution, only the single PL model with $\alpha_{\rm SW}$ is considered.

For the transport of CRPs inside the SBN, we have adopted two different diffusion models; CRPs scatter off either the large-scale turbulence of $\delta B_{\rm SBN}/B_{\rm SBN} \sim 1$ present in the SBN (diffusion model A) or the waves self-excited by CR streaming instability (diffusion model B) \citep{peretti2019,krumholz2020}. 
The diffusion model B assumes that the pre-existing turbulence of scales less than the gyroradius of CRPs with $E_{\rm CR} \lesssim {\rm several} \times 0.1~{\rm PeV}$ damps out by ion-neutral collisions. 

Using the CRP momentum spectrum given in Equation (\ref{eq9}) and the analytic approximations for source functions presented by \citet{kelner2006}, 
we have then calculated the $\gamma$-ray and neutrino emissions due to $pp$ collisions in the SBN,
and estimated their fluxes from three nearby SBGs, M82, NGC253, and Arp220, observable at the Earth.

Main findings are summarized as follows.

(1) If the single PL model for SNR-produced CRPs is adopted, the $\gamma$-ray observations of Fermi-LAT, Veritas, and H.E.S.S. are better reproduced with the diffusion model A than with the model B. If the double PL model is adopted, on the other hand, the observations seem to be better reproduced with the diffusion model B. However, other combinations, although statistically somewhat less significant, cannot be excluded.

(2) The contribution of SW-produced CRPs may depend on the IGIMF; SBGs with higher SFRs are likely to have
flatter IGIMFs, resulting in larger ratios of $\mathcal{L}_{\rm SW}/\mathcal{L}_{\rm SN}$. In Arp220 where the star formation rate is high, the SW-produced CRPs may make a significant contribution to the $\gamma$-ray emission.

(3) The predicted neutrino fluxes from three nearby SBGs seem too small, so that SBGs would not be observed as point sources by IceCube. On the other hand, our estimation suggests that M82 and NGC253 might be detectable as point sources with the upcoming KM3NeT and IceCube-Gen2, if the diffusion model A along with $p_{\rm max}=100$ PeV/$c$ is employed. In other models, the predicted neutrino fluxes would not be high enough to be detected as point sources by KM3NeT or IceCube-Gen2.

(4) Considering uncertainties in the CRP production and transport models in the SBN, we have examined various combinations of models in this paper. Based on our results, we point that currently available $\gamma$-ray observations may not provide sufficient information to differentiate the CRP production and transport models. 
However, those models could be further constrained with ``multi-messenger observations'' including possible detections of neutrinos from SBGs with future neutrino telescopes, such as KM3NET and IceCube-Gen 2, as well as further TeV-range $\gamma$-rays observations.

Finally, \citet{muller2020} recently suggested that high-energy processes associated with superwinds driven by the supersonic outflows from the SBN (not stellar winds) can produce CRPs as well. But their contribution to $\gamma$-ray emissions from SBGs would be $\lesssim10\%$, so that the main results of this paper would not be changed even when such contribution is included. 

\acknowledgments
We thank the anonymous referee for constructive comments that help us improve this paper from its initial form. We also thank Dr. K. Murase for comments on the manuscript. J.-H.H. and D.R. were supported by the National Research Foundation (NRF) of Korea through grants 2016R1A5A1013277 and 2020R1A2C2102800. J.-H. H. was also supported by the Global PhD Fellowship of the NRF through grant 2017H1A2A1042370. H.K. was supported by the Basic Science Research Program of the NRF through grant 2020R1F1A104818911.

\bibliography{StarBurstG}{}
\bibliographystyle{aasjournal}

\end{document}